\documentclass[12pt]{article}
\usepackage{amssymb,amsmath,amsfonts,amsthm,amstext,amscd,array}
\usepackage{mathrsfs}
\usepackage{hyperref}
\usepackage{pdfsync}
\usepackage{bbm}
\usepackage{bm}
\usepackage[arrow,matrix,curve]{xy}
\usepackage{bbding}
\usepackage{wasysym}

\usepackage{color,soul}

\usepackage{amsfonts}
\usepackage{booktabs}

\usepackage{epsf}
\usepackage{epsfig}
\allowdisplaybreaks

\newcommand{\dd}{{\mathsf{d}}}

\def\dd{{\rm d}}

\newcommand{\eq}{\begin{equation}}
\newcommand{\eqend}{\end{equation}}
\newcommand{\eqa}{\begin{eqnarray}}
\newcommand{\nonueqa}{\begin{eqnarray*}}
\newcommand{\eqaend}{\end{eqnarray}}
\newcommand{\nonueqaend}{\end{eqnarray*}}

\newcommand{\bma}[1]{\begin{array}{#1}}
\newcommand{\ema}{\end{array}}
\newcommand{\bc}{\begin{center}}
\newcommand{\ec}{\end{center}}

\newif\ifold             \oldtrue

\def\be{\begin{equation}}
\def\ee{\end{equation}}
\def\bea{\begin{eqnarray}}
\def\eea{\end{eqnarray}}
\def\bd{\begin{displaymath}}
\def\ed{\end{displaymath}}

\newcommand{\beq}{\begin{eqnarray}}
\newcommand{\eeq}{\end{eqnarray}}

\makeatletter
\newdimen\normalarrayskip              
\newdimen\minarrayskip                 
\normalarrayskip\baselineskip
\minarrayskip\jot
\newif\ifold             \oldtrue            
\def\arraymode{\ifold\relax\else\displaystyle\fi} 
\def\@arrayskip{\ifold\baselineskip\z@\lineskip\z@
     \else
     \baselineskip\minarrayskip\lineskip2\minarrayskip\fi}
\def\@arrayclassz{\ifcase \@lastchclass \@acolampacol \or
\@ampacol \or \or \or \@addamp \or
   \@acolampacol \or \@firstampfalse \@acol \fi
\edef\@preamble{\@preamble
  \ifcase \@chnum
     \hfil$\relax\arraymode\@sharp$\hfil
     \or $\relax\arraymode\@sharp$\hfil
     \or \hfil$\relax\arraymode\@sharp$\fi}}
\def\@array[#1]#2{\setbox\@arstrutbox=\hbox{\vrule
     height\arraystretch \ht\strutbox
     depth\arraystretch \dp\strutbox
     width\z@}\@mkpream{#2}\edef\@preamble{\halign \noexpand\@halignto
\bgroup \tabskip\z@ \@arstrut \@preamble \tabskip\z@ \cr}%
\let\@startpbox\@@startpbox \let\@endpbox\@@endpbox
  \if #1t\vtop \else \if#1b\vbox \else \vcenter \fi\fi
  \bgroup \let\par\relax
  \let\@sharp##\let\protect\relax
  \@arrayskip\@preamble}
\makeatother

\newcommand{\p}{\partial}

\def\be{\beta}

\newtheorem{assumption}{Assumption}[section]

\newtheorem{definition}[assumption]{Definition}

\theoremstyle{definition}

\usepackage{color,xcolor}
\usepackage{manfnt}

\def\p{\vskip2ex\hskip5ex}
\def\ddo{\end{document}}
\usepackage{hyperref}
\hypersetup{colorlinks,%
citecolor=black,%
filecolor=black,%
linkcolor=black,%
urlcolor=black}

\newcommand{\nc}{\newcommand}
\nc{\lb}{\llbracket}
\nc{\rb}{\rrbracket}
\nc{\gl}{\llbracket}
\nc{\gr}{\rrbracket}

\begin{document}

\title{\Large\bf  Non-commutative gauge symmetry from strong homotopy algebras}
\author{{Vladislav Kupriyanov$^1$\thanks{vladislav.kupriyanov@gmail.com}, Fernando Oliveira$^1$\thanks{fernando.martins@ufabc.edu.br}, Alexey Sharapov$^{1,2}$\thanks{sharapov@phys.tsu.ru}}\\{ and Dmitri Vassilevich$^1$\thanks{dvassil@gmail.com}}\\ \\{$^1$ \it Centro de Matemática, Computação e Cognição - Universidade Federal do ABC,}\\{\it  Santo André, SP, Brazil}\\{$^2$}{\it Physics Faculty, Tomsk State University,  Lenin ave. 36, Tomsk 634050, Russia}}
\maketitle

\p
\begin{abstract}
We explicitly construct an L$_\infty$ algebra that defines U$_{\star}(1)$ gauge transformations on a space with an arbitrary noncommutative and even nonassociative star product. Matter fields are naturally incorporated in this scheme as  L$_\infty$ modules. Some possibilities for including  P$_\infty$  algebras are also discussed. 
\end{abstract}

\section{Introduction}\label{sec:int}
Constructing gauge theories on noncommutative spaces is a notoriously difficult problem. There are many proposals and schemes on the market, which have their advantages and disadvantages \cite{Hersent:2022gry}. To us, the  L$_\infty$ approach to gauge theories \cite{Hohm:2017pnh, Zeitlin:2007vv} applied to noncommutative spaces in \cite{Blumenhagen:2018kwq, Blumenhagen:2018shf} seems to be the most promising. However, there is a caveat: so far there is no proof that desired L$_\infty$ algebras exist to all orders in the parameter of noncommutativity. Filling this gap is the principal goal of the present work.

In this paper, we do not restrict ourselves to associative noncommutative spaces. Nonassociative star products appear in the context of magnetic backgrounds in field theory \cite{Jackiw:1984rd}, non-geometric fluxes \cite{Lust:2010iy, Blumenhagen:2010hj, Mylonas:2012pg} and D-branes \cite{Schomerus:1999ug, Cornalba:2001sm, Herbst:2003we} in string theory, see \cite{Szabo:2019hhg} for a review. To the readers who are not interested in nonassociative geometries, we mention that associativity does not bring drastic simplifications to our L$_\infty$ construction.

The strong homotopy Lie algebras, or the L$_\infty$ algebras, are being used in many areas of theoretical physics and are especially efficient in the treatment of quantization problems (see \cite{Lada:1992wc} for an introduction). In contrast to the Lie algebras, the L$_\infty$ algebras are defined through infinitely many $n$-linear maps $\ell_n$ on a $\mathbb{Z}$-graded vector space $V$. The maps are defined to satisfy an infinite sequence of relations called the generalized Jacobi identities. In the present work, we are interested in a noncommutative U$(1)$ gauge theory. Namely, we define gauge transformations of the gauge connections. Therefore we take $V=V_0\oplus V_1$ with $V_0$ containing zero-forms identified with the gauge parameters and $V_1$ containing gauge field one-forms. The map $\ell_2$ on $V_0\otimes V_0$ is defined by the star product and used as an input. Other maps are determined (non-uniquely) through the generalized Jacobi identities. The procedure of solving the generalized  Jacobi identities (which is called the L$_\infty$ bootstrap) is organized as a sequence of steps. At each step, we solve two equations. The construction exploits the standard homological perturbation theory\footnote{Equivalent results may be also achieved  by solving directly the recurrence relations similarly to the construction of star products in \cite{Kupriyanov:2008dn, Kupriyanov:2015dda}. This latter method is much more technically demanding. We do not present any details here.}. We  would like to stress that in this way we construct noncommutative U$(1)$ gauge transformations for an arbitrary star product.
In this setting, it is natural to interpret the matter fields as an L$_\infty$ module. The construction of such a module is completely similar to the construction of the L$_\infty$ algebra itself.

Since we allow for nonassociative star products we also allow for almost Poisson structures which do not need to be Lie--Poisson meaning that the corresponding bivector does not need to satisfy the Jacobi identity. In this context, it is natural to ask the following question.
Which algebraic condition may replace the associativity in the quantization of almost Poisson structures which are not Lie--Poisson? The possibilities of imposing identities in the deformed algebra have been considerably reduced by the no-go results \cite{Bojowald:2016lnl,Kupriyanov:2017oob,Vassilevich:2018gkl} and practically closed by the recent work \cite{Dotsenko:2023lvq}. A promising algebraic framework seems to be provided by the strong homotopy associative algebras A$_\infty$ \cite{Stasheff:1963a, Stasheff:1963b}. The relation between these algebras and the L$_\infty$ algebras is similar to that between the associative and Lie algebras. The A$_\infty$ algebras describe a controlled (up to homotopy) violation of associativity. In contrast to the L$_\infty$ algebras the multiplication maps $m_n$ of A$_\infty$ algebras do not have any prescribed symmetry properties, which makes the analysis of corresponding homotopy relations much more complicated. A counterpart of A$_\infty$ algebras in the Poisson setting is given by the strong homotopy 
Poisson algebras  introduced by Cattaneo and Felder \cite{Cattaneo:2007}. In the present work, we attempt to construct P$_\infty$ algebras starting with an almost Poisson structure with two different choices for a graded commutative product on $V$ (which is another ingredient of P$_\infty$ algebras). We show that in both  cases, the P$_\infty$ relations imply the Jacobi identity for the almost Poisson bracket.

The paper is organized as follows. In the next Section, we present  the main definitions and explain the  logic behind our construction of L$_\infty$ algebras. The construction itself is contained in Section \ref{sec:constr}. In Section \ref{sec:module}, we introduce L$_\infty$ modules associated with  matter fields and prove their existence.  In Section \ref{sec:APinfty}, we discuss the A$_\infty$ and P$_\infty$ algebras.
Section \ref{sec:concl} contains some concluding remarks.

\section{Setup and main definitions}\label{sec:set}
 Let $V=\bigoplus_{k\in  \mathbb{Z}}\, V_{k}$ be a $ \mathbb{Z}$-graded vector space and let $|v|=k$ denote the degree of a 
 homogeneous element $v\in V_k$.  By definition, an L$_\infty$-algebra is the vector space $V=\bigoplus_{k\in  \mathbb{Z}}\, V_{k}$ together with a sequence of graded-antisymmetric multilinear maps $
\ell_n:V^{\otimes n}\to V$ of  degree $|\ell_n|=2-n$. The graded antisymmetry means that
\begin{equation}
\ell_n(\dots,v_j,v_{j+1},\dots)=-(-1)^{|v_j|\cdot |v_{j+1}|} \ell_n(\dots,v_{j+1},v_{j},\dots). \label{gradant}
\end{equation}
The maps should satisfy the
identities ${\cal J}_n(v_1,\dots,v_n)=0$ for each $n\geq1$, called the generalized Jacobi identities, with
\begin{eqnarray} \label{eq:calJndef}
{\cal J}_n(v_1,\dots,v_n) &:=& \sum_{i+j=n+1}\, (-1)^{i\,(n-i)} \
  \sum_{\sigma\in{\rm Sh}_{i,n-i}} \,  \chi(\sigma;v_1,\dots,v_n) \\ &&
                                                                      \qquad
                                                                \times
                                                                 \ell_{j}\big(
                                                             \ell_i(v_{\sigma(1)},\dots,
                                                              v_{\sigma(i)}),
                                                              v_{\sigma(i+1)},\dots
                                                              ,v_{\sigma(n)}\big)
                                                                      \
                                                        \notag\,.
\end{eqnarray}
Here, the second sum runs over $(i,n-i)$-shuffled permutations
$\sigma\in S_n$ of degree $n$ which are restricted as 
\begin{equation*}
\sigma(1)<\dots<\sigma(i) \,, \qquad \sigma(i+1)<\dots < \sigma(n) 
\end{equation*}
and $\chi(\sigma;v_1,\dots,v_n)=\pm\, 1$ is the Koszul sign defined by the relation
\begin{equation}\ell_n(v_1,\dots,v_n)=\chi(\sigma;v_1,\dots,v_n)\, \ell_n(v_{\sigma(1)},\dots,v_{\sigma(n)})\,.
\end{equation}
This completes the definition of generic L$_\infty$ algebras. 

We will be interested only in the so-called flat L$_\infty$ algebras for which 
$
\ell_0=0
$.
Later we will impose some further restrictions on the class of such algebras.

The arena for our construction is a formal quantization of $\mathbb{R}^p$ in the direction of an almost Poisson structure  defined by a smooth bivector $P$ on $\mathbb{R}^p$. This bivector in turn defines an almost Poisson bracket
\begin{equation}
\{ f,g\} =P(\dd f,\dd g)=P^{ij}\partial_i f \cdot \partial_jg \label{almP}
\end{equation}
for $f,\, g\in C^\infty (\mathbb{R}^p)$. A formal noncommutative (and nonassociative) structure on $\mathbb{R}^p$ is defined through a star product.
Let $\mathcal{A}=C^\infty (\mathbb{R}^p)[[\lambda]]$ be an algebra of formal power series in the parameter $\lambda$ with coefficients in $C^\infty (\mathbb{R}^p)$. A star product is a deformation of the pointwise product on $\mathcal{A}$. It is a bilinear map $\star: \mathcal{A}\times\mathcal{A}\to \mathcal{A}$ of the form
\begin{equation}
f\star g = f \cdot g +\sum_{r=1}^\infty \lambda^r C_r(f,g) \,,\label{starex}
\end{equation}
where the $C_r$'s are bidifferential operators and
\begin{equation}
C_1(f,g)=\{ f, g\} \,. \label{C1}
\end{equation}
We shall consider exclusively the star products which satisfy the stability of unity condition
\begin{equation}
1\star f = f\star 1=f \label{stabun}
\end{equation}
for all $f\in C^\infty (\mathbb{R}^p)[[\lambda]]$. We do not impose any further conditions on the star product. In particular, we do \emph{not} assume that this product is associative. Thus, we do not need to assume that the bivector $P$ satisfies the  Jacobi identity.

Our purpose is to describe U$(1)$ gauge transformations on noncommutative spaces. The minimal requirement is to have gauge parameters (which are 0-forms) and gauge fields (which are 1-forms). Therefore, we take $V_0=\Omega^0 (\mathbb{R}^p)[[\lambda]]$, $V_1=\Omega^1 (\mathbb{R}^p)[[\lambda]]$, and set $V_k=\{ 0\}$ for $k\neq 0,1$. Gauge parameters are denoted by $f, g, h \in V_0$ and gauge fields by $A, B \in V_1$.

The star product enters the game through our choice of the bilinear map $\ell_2$ on $V_0$:
\begin{equation}
 \ell_2(f,g)=[f,g]_\star \equiv f\star g - g\star f\,. \label{b1}
\end{equation}
In this formula, only the antisymmetric part of the star appears. The complete product will appear in Section \ref{sec:module} where we will study the matter fields.
The stability of unity condition (\ref{stabun}) yields $\ell_2(1,f)=0$. We also require that
\begin{equation}
\ell_2(1,A)=0\label{stabunA}
\end{equation}
for all $A\in V_1$. Then the condition ${\cal J}_2(1,f)=0$ gives
\begin{equation}
\ell_2(\ell_1(1),f)=0,
\end{equation}
which is satisfied by
\begin{equation}
\ell_1(g)=\dd g,\label{ell1} 
\end{equation}
where $\dd$ is the de Rham differential extended to formal series by linearity. 

By counting the degrees, we obtain that non-zero multilinear maps $\ell_n$ may contain either one scalar and $n-1$ one-forms or two scalars and $n-2$ one forms. Similarly, non-trivial conditions ${\cal J}_n$ may contain either two or three scalars. The L$_\infty$ relations will be solved according to the following scheme,
\begin{center}
\begin{tabular}{c| c| c} \toprule
    {\mbox{Step}} & {L$_\infty$ relations to solve} & {$\ell$-maps to determine}  \\ \midrule
    {1-st}& {${\cal J}_2(f,g)=0$}& {$\ell_2(f,A)$}\\
    &{${\cal J}_3(f,g,h)=0$}& {$\ell_3(f,g,A)$}\\  \midrule
    {2-nd}& {${\cal J}_3(f,g,A)=0$}& {$\ell_3(f,A,A)$} \\
     & {${\cal J}_4(f,g,h,A)=0$}& {$\ell_4(f,g,A,A)$} \\  \midrule
       {3-rd}&  {${\cal J}_4(f,g,A,A)=0$}& {$\ell_4(f,A,A,A)$} \\
     & {${\cal J}_5(f,g,h,A,A)=0$}& {$\ell_5(f,g,A,A,A)$} \\   \midrule
     {$\dots$}& {$\dots$} & {$\dots$}
     \\ \bottomrule
\end{tabular}
\end{center} 
For example, in the first step, we will use the relation ${\cal J}_2(f,g)=0$ to determine $\ell_2(f, A)$ and the relation ${\cal J}_3(f,g,h)=0$ to define $\ell_3(f,A,A)$. We call this iterative procedure the L$_\infty$ bootstrap. 

Some comments are in order. First, the solutions for $\ell_n$ will be highly non-unique and contain inevitable ambiguities (see the very end of Sec. \ref{Sec32}).  Second, it will be enough to consider only the ``diagonal'' elements of the multilinear maps computed for coinciding arguments from $V_1$. These elements are in one-to-one correspondence with non-diagonal elements. Indeed, for any symmetric $n$-linear map $\mathcal{L}_n$ one has the identity
\begin{equation}\label{polar}
\mathcal{L}_n(A_1,\dots,A_n)=\frac 1{n!} {\partial_{z^1}\cdots \partial_{z^n}}\, \mathcal{L}_n(\hat A,\dots,\hat{A}),
\end{equation}
where $\hat{A}=z^1A_1+\dots +z^n A_n$ with real variables $z^j$. Furthermore, when evaluating the  $\ell_n$'s or $\mathcal{J}_n$'s on $V^{\otimes n}$, we can restrict ourselves to expressions  in which the arguments from $V_0$ precede those from $V_1$, as in the table above. The other distributions of arguments are obtained  by graded antisymmetry (\ref{gradant}). 

Once the L$_\infty$ algebra has been constructed, the gauge transformations can be defined as \cite{Hohm:2017pnh}
\begin{eqnarray}\label{q2}
  \delta_{f}  A :=\dd f-[\![ f,A]\!]_A=\sum^\infty_{n= 0}   \frac1{n!}
      (-1)^{\frac{n(n-1)}{2}}\,
 \ell_{n+1}(f, A, \dots, A )\,.
\end{eqnarray}
The generalized Jacobi identities with two gauge parameters, ${\cal J}_{n+2}(f,g,A^{\otimes n})=0$, imply that
\begin{eqnarray}\label{fncga}
[\delta_{f},\delta_{g}]A = \delta_{[\![f,g]\!]_A} A\,,\end{eqnarray}
where
\begin{eqnarray}\label{q1}
[\![f,g]\!]_A: = -\sum_{n=0}^\infty \, \frac1{n!}\,
 (-1)^{\frac{n(n-1)} {2}}\, \ell_{n+2}\left(f,g,A^{\otimes n} \right)\,.
\end{eqnarray}
In addition, the identities with three gauge parameters, ${\cal J}_{n+3}(f,g,h,A^{\otimes n})=0$, imply the relation \cite{Kupriyanov:2021cws}
\begin{equation}\label{fgh}
[\![h,[\![f,g]\!]_A]\!]_A+\delta_h[\![f,g]\!]_A+\mbox{cycl}(f,g,h)=0\,, 
\end{equation}
which in turn guarantees the Jacobi identity for gauge variations $\delta_f$,
 \begin{equation}
 [\delta_f,[\delta_g,\delta_h]]+\mbox{cycl}\equiv0\,.\label{Jacobi}
 \end{equation}
In the zero order in $A$, one recovers the usual formulas 
\begin{eqnarray}\label{ic3}
\delta_f(A)=\dd f+{\cal O}(A),\qquad [\![f,g]\!]_A&=&-[f,g]_\star+{\cal O}(A)\,,
\end{eqnarray}
which are shared by most of the approaches to noncommutative U$(1)$ gauge theories.

\section{Construction of the L$_\infty$ algebra}\label{sec:constr}

In this section, we apply the method of L$_\infty$ bootstrap \cite{Blumenhagen:2018kwq, Blumenhagen:2018shf} to construct an L$_\infty$ algebra underlying the noncommutative gauge symmetry. We show that, under mild assumptions, the equations of L$_\infty$ bootstrap can be solved explicitly up to any given order in the  gauge field $A$. 

\subsection{The first step}\label{sec:first}
At the first step of L$_\infty$ bootstrap, we define $\ell_2(f,A)$ and $\ell_3(f,g,A)$ which solve the conditions ${\cal J}_2(f,g)=0$ and ${\cal J}_3(f,g,A)=0$, respectively. 

The relation ${\cal J}_2(f,g)=0$ reads
\begin{equation}
\ell_2\left(\ell_1(f),g\right)+\ell_2\left(f,\ell_1(g)\right)=\ell_1\left(\ell_2(f,g)\right)\,.\label{b2}
\end{equation}
We can write the expression on the right as
\begin{equation}\label{b7}
\ell_1\left(\ell_2(f,g)\right)=\dd \ell_2(f,g)=\sum_{p,q=1}^\infty G_a^{(i)^p(j)^q}\left(\partial_i\right)^pf\left(\partial_j\right)^qg\,\dd x^a,
\end{equation}
where the structure functions of the bidifferential operator $\ell_1\circ \ell_2$ satisfy the symmetry property 
\begin{equation}
G_a^{(i)^p(j)^q}=-G_a^{(j)^q(i)^p}\,. \label{b8}
\end{equation}
Hereinafter we use the multi-index notation
\begin{equation}\label{mi}
(i)^p=(i_1\dots i_p)\,,\qquad \left(\partial_i\right)^p=\partial_{i_1}\dots\partial_{i_p}\,.
\end{equation}
If a multi-index is included in round brackets, this means symmetrization. 
Due to the stability of unity condition, $\ell_2(f,g)$ contains no terms without derivatives of $f$ or $g$. Therefore, the summation in (\ref{b7}) starts with $p,q=1$. 

Now, one easily finds the following  solution to (\ref{b2}):
\begin{equation}\label{b3}
\ell_2(f,A)=\frac 12 \sum_{p,q=1}^\infty G_a^{(i)^p(j)^{q}}\left(\partial_i\right)^pf\left(\partial_j\right)^{q-1}A_{j_q}\,\dd x^a.
\end{equation}

To determine $\ell_2(f,g,A)\in V_0$ we need to solve the equation
\begin{eqnarray}\label{b4}
&\ell_3(\ell_1(f),g,h)+\ell_3(f,\ell_1(g),h)+\ell_3(f,g,\ell_1(h))=\\
&-\ell_2(\ell_2(f,g),h)-\ell_2(\ell_2(g,h),f)-\ell_2(\ell_2(h,f),g)\,.\notag
\end{eqnarray}
The right hand side is clearly antisymmetric in $f$, $g$ and $h$ and we denote it by 
\begin{equation}
\Pi_\star(f,g,h)=[[f,g]_\star,h]_\star+[[g,h]_\star,f]_\star+[[h,f]_\star,g]_\star\,.
\end{equation}
Moreover, the stability of unity implies that
\begin{equation}
\Pi_\star(f,g,h)=\hat\Pi_\star(\dd f,\dd g,\dd h)\,,
\end{equation}
where the polydifferential operator 
\begin{eqnarray}\label{c2}
\hat\Pi_\star(\dd f,\dd g,\dd h)=\sum_{p,q,r=1}^\infty F^{(i)^p(j)^q(k)^{r}}\left(\partial_i\right)^pf\left(\partial_j\right)^qg\left(\partial_k\right)^rh
\end{eqnarray}
 is antisymmetric with respect to the permutation of $f$, $g$ and $h$, so the coefficients $F^{(i)^p(j)^q(k)^{r}}$ are 
 antisymmetric with respect to the permutations of the groups of indices. 
Now, one can see that the expression
\begin{equation}\label{b5}
\ell_3(f,g,A)=\frac13 \hat\Pi_\star(\dd f,\dd g, A)\end{equation}
satisfies Eq. (\ref{b4}). This completes the first step of induction.

\subsection{All order solution}\label{Sec32}
At the $(m+1)$-th step of the L$_\infty$ bootstrap, we are trying to determine the structure maps 
\begin{equation}\label{sm}
\ell_{m+2}(f,A^1,\ldots, A^{m+1})\quad \mbox{and}\quad \ell_{m+3}(f,g,A^1,\ldots, A^{m+1})
\end{equation}
from the generalized Jacobi identities 
\begin{eqnarray}\label{JJ}
{\cal J}_{m+2}\left(f,g,A^1,\dots,A^m\right)=0 \quad \mbox{and}\quad
{\cal J}_{m+3}\left(f,g,h,A^1,\dots,A^m\right)=0\,.
\end{eqnarray}

The complexity of the analysis increases rapidly with $m$. Therefore, to proceed further, we need some preparation. First of all, it is convenient to replace scalar parameters 
\begin{equation}\label{fff}
f_1,f_2,f_3,\ldots\in C^{\infty}(\mathbb{R}^p)
\end{equation}
by a single function $C$ on $\mathbb{R}^p$. To this end, let us introduce the  Grassmann algebra $\mathrm{Gr}_m$ on $m$ anti-commuting generators $\xi_1,\dots,\xi_m$ and define $C=\xi_1f_1+\xi_2f_2+\cdots +\xi_mf_m$. Then any skew-symmetric function $\mathcal{A}_n(f_1,f_2,\dots,f_n)\in C^{\infty}(\mathbb{R}^p)$ with $n\leq m$ defines by $\mathrm{Gr}_m$-linearity an element $\mathcal{A}_n(C,\dots,C)\in C^{\infty}(\mathbb{R}^p)\otimes \mathrm{Gr}_m$. The original function is restored by the relation
\begin{equation}
    \mathcal{A}_n(f_1,\dots,f_n)=\frac 1{n!}\partial_{\xi_n}\cdots \partial_{\xi_1} \, \mathcal{A}_{m}(C,\dots,C)|_{\xi=0}\,.
    \label{polar2}
\end{equation}
This formula is an odd counterpart of (\ref{polar}). All Grassmann derivatives are left derivatives, so that, e.g., $\partial_{\xi_2}\partial_{\xi_1}\xi_1\xi_2=1$. In this work, we will never use any relations with more than four arguments in $V_0$. Thus, we may take $m=4$. The action of the exterior differential $\dd$ also extends from (\ref{fff}) to $C$ by $\mathrm{Gr}_m$-linearity.

With the definitions above the structure maps (\ref{sm}) are completely determined by the following homogeneous functions of $C$ and $A$:
\begin{equation}\label{cca}
\ell_{m+2}(C,A,\ldots, A)\quad \mbox{and}\quad \ell_{m+3}(C,C,A,\ldots, A)\,.
\end{equation}
As a next step, we isolate the terms involving $\ell_1 (C)=\dd C$  and rewrite equations (\ref{JJ}) as\footnote{The superscript $R$ refers to the `remainder' of the generalized Jacobi identities (\ref{JJ}). }
\begin{equation}\label{hpt}
\begin{array}{rcl}
(\delta \ell_{m+2})(C,C; A,\ldots, A)&=&{\cal J}^R_{m+2}\left(C,C;A,\ldots, A\right)\,,\\[3mm]
(\delta \ell_{m+3})(C,C, C;A,\ldots, A)&=&{\cal J}^R_{m+3}\left(C,C,C;A,\ldots, A\right)\,.
\end{array}
\end{equation}
The operator $\delta$ acts in the space of functions (\ref{cca}) by the general rule:
\begin{equation}\label{delta}
    (\delta \mathcal{L})(\underbrace{C,\ldots,C}_{p+1}; \underbrace{A,\ldots, A}_{q-1})=q\mathcal{L}(C,\ldots, C; \dd C, A,\ldots, A )
\end{equation}
for any homogeneous function
\begin{equation}\label{coch}
    \mathcal{L}(\underbrace{C,\ldots, C}_{p};\underbrace{A,\ldots,A}_{q})\,.
\end{equation}
The right-hand sides of equations (\ref{hpt}) are determined by the functions

\begin{equation}\label{JnfgA1}    
\begin{array}{c}
{\cal J}^R_{m+2}(f,g,A,\ldots, A)\\[5mm]
\displaystyle =\sum_{i+j=m-1}^{}(-1)^{i\,j+n}\frac{m!}{(i+1)!\,j!}\Big[(-1)^j\,\ell_{i+2}\left(\ell_{j+2}(f,g,A^{\otimes j}),A^{\otimes i+1}\right)
\\[7mm]
\displaystyle -\ell_{j+2}\left(\ell_{i+2}(f,A^{\otimes i+1}),g,A^{\otimes j}\right)+\ell_{j+2}\left(\ell_{i+2}(g,A^{\otimes i+1}),f,A^{\otimes j}\right)\Big] \\[5mm]
\displaystyle+(-1)^{m}\ell_1\left(\ell_{n+2}(f,g,A^{\otimes m})\right)
\end{array}
\end{equation}
and
\begin{equation}\label{JnfgA2}    
\begin{array}{c}
{\cal J}^R_{m+3}(f,g,h,A,\ldots, A)\\[5mm]
\displaystyle =(-1)^{m+1}\sum_{i+j=m-1}(-1)^{i\,j}\frac{m!}{i!\,(j+1)!}\Big[\ell_{i+3}\left(\ell_{j+2}(f,A^{\otimes j+1}),g,h,A^{\otimes i}\right) \\[5mm]
+(-1)^{i}\ell_{j+3}\left(\ell_{i+2}(f,g,A^{\otimes i}),h,A^{\otimes j+1}\right)\Big] \\[5mm]
+(-1)^{n+1}\ell_{2}\left(\ell_{m+2}(f,g,A^{\otimes n}),h\right)+\mbox{cycl.}\big(f,g,h\big)\,.
\end{array}
\end{equation}
These are given by various compositions of $\ell_{n+2}(f, A^{\otimes(n+1)})$ and $\ell_{n+3}(f,g,A^{\otimes(n+1)})$ with $n<m$ that have been defined in previous stages.   It is straightforward to see that the operator (\ref{delta}) squares to zero, making the space of functions (\ref{coch}) into a cochain complex. System (\ref{hpt}) assumes thus the standard form of homological perturbation theory: $\delta x_{n+1}=f_n(x_1,\ldots, x_n) $ where the differential $\delta$ and the functions $f_n$ are given, and the $x_{n+1}$'s are unknowns \footnote{For more details on homological perturbation theory we refer the reader to \cite[Ch. 8.4]{henneaux1994quantization}. Some other technical tools can be found in \cite{JRSM}.}. The solubility of such systems is controlled by the cohomology of the coboundary operator $\delta$.  Applying $\delta$ to both sides of (\ref{hpt}) yields the cocycle conditions 
\begin{equation}
(\delta {\cal J}^R_{m+2})(C,C,C;A,\ldots,A)=0 \quad \mbox{and}\quad (\delta {\cal J}^R_{m+3})(C,C,C,C;A,\ldots,A)=0\,.
\end{equation}
These conditions are necessary for the equations  to have a solution. Sufficiency requires more: both the cocycles must be trivial. Direct calculation shows that the right-hand sides of equations (\ref{hpt}) are indeed $\delta$-closed, provided that all previous equations are satisfied (see \cite{KupDurham} for the first non-trivial orders). If the differential were acyclic, this would imply that equations  (\ref{hpt}) are solvable. However, this is not the case.  For instance, the functions
\begin{equation}
\mathcal{L}(C)=C\,,\qquad \mathcal{L}(A,A)=F^2 \,, \qquad \mathcal{L}(C,A,A)=CF^2\,,
\end{equation}
where $F_{ij}\dd x^i\wedge \dd x^j=\dd A$ and $F^2=F_{ij}F^{ij}$, are nontrivial $\delta$-cocycles and one can easily construct more examples. 
To avoid possible obstructions to solvability we  impose certain restrictions on the structure maps of the $L_\infty$-algebra to be found. First, we are looking for the $\ell_n$'s that are polydifferential operators with coefficients in $\Omega^0(\mathbb{R}^p)\oplus \Omega^1(\mathbb{R}^p)$. Second, the polydifferential operators  should respect the unit, meaning that  both $\ell_{m+1}(f, A^{\otimes m})$ and $\ell_{m+2}(f,g, A^{\otimes m})$ must vanish whenever one of their arguments $f$ and $g$ is equal to $1$. We will refer to such polydifferential operators as {\it unital}. It is significant that the polydifferential operators defining the generalized Jacobi identities (\ref{JJ}) are always unital if constructed from unital $\ell_n$'s and the same is true for the right-hand sides of equations (\ref{hpt}). 

Let us now introduce the following infinite sets of fields
\begin{equation}\label{zuw}
\begin{array}{c}
    \{z^\alpha\}:=\big\{\partial_{i_1}\ldots\partial_{i_{n+1}}C\big\}_{n=0}^\infty\,,\qquad \{u^\alpha\}=\big \{ \partial_{(i_1}\ldots\partial_{i_n} A_{i_{n+1})} \big\}_{n=0}^\infty\,,\\[5mm]
    \{w^J\}=\big\{ \partial_{(i_1}\ldots\partial_{i_n}F_{i)j}\big\}_{n=0}^\infty\,.
    \end{array}
\end{equation}
Here the index $\alpha$ runs over the multi-indices $(i)^n$, $n=1,2,\ldots$, and $J$ runs over $(i)^n j$, $n=1,2,\ldots$.  It is easy to see that  any unital polydifferential operator  evaluated on $A$ and $C$ can be written uniquely as a polynomial in $z$'s, $u$'s, and $w$'s, 
\begin{equation}\label{pol}
     \mathcal{L}(\underbrace{C,\ldots, C}_{p};\underbrace{A,\ldots,A}_{q})=\sum_{k+l=q} f_{\alpha_1\cdots \alpha_p\beta_1\cdots \beta_k J_1\cdots J_l}z^{\alpha_1}\cdots z^{\alpha_p}u^{\beta_1}\cdots u^{\beta_k}w^{J_{1}}\cdots w^{J_l}\,,
\end{equation}
where  only a finite  number of the coefficients $f$ are different from zero. 
In this notation, the action of the differential (\ref{delta}) is given by the formula
\begin{equation}
    \delta\mathcal{L}=z^\alpha\frac{\partial\mathcal{L} }{\partial u^\alpha} =\sum_{n=1}^\infty\partial_{i_1}\cdots\partial_{i_n}C\frac{\partial\mathcal{L}}{\partial (\partial_{i_1}\cdots\partial_ {i_{n-1}}A_{i_n})}\,.
\end{equation}
In order to show the acyclicity of $\delta$ in the space of unital polydifferential operators (\ref{coch}) with $p>0$, we introduce the operator 
\begin{equation}
    \delta^\ast\mathcal{L}=u^\alpha\frac{\partial\mathcal{L}}{\partial z^\alpha}=\sum_{n=1}^\infty\partial_{i_1}\cdots\partial_{ i_{n-1}}A_{i_n}\frac{\partial\mathcal{L}}{\partial (\partial_{i_1}\cdots\partial_{i_n}C)}\,.
\end{equation}
Clearly,
\begin{equation}\label{N}
    \delta\delta^\ast+\delta^\ast\delta=N\,,\qquad  N=z^\alpha\frac{\partial}{\partial 
    z^\alpha}+u^\alpha\frac{\partial}{\partial u^\alpha}\,.
\end{equation}
The operator $N$ counts the total degree of a polynomial in the variables $z^\alpha$ and $u^\alpha$. In particular,  $N$ is invertible on the subspace of polynomials (\ref{pol}) at least linear in $z$'s. Upon restricting to this subspace, we can write
\begin{equation}
    (N^{-1}f)(z,u,w)=\int_0^1 f(tz,tu, w)\frac{dt}{t}\,.
\end{equation}
With the commutation relation (\ref{N}), one can easily see that the expressions\footnote{  The operator $h=\delta^\ast N^{-1}=N^{-1}\delta^\ast$ is the standard contracting homotopy for $\delta$ often used in local BRST cohomology, see e.g. \cite[App. A]{Barnich_2004}. It is also obtained by the symmetrization of the homotopy operator of \cite{Kupriyanov_2023}.}
\begin{equation}\label{rr}
\begin{array}{rcl}
\ell_{m+2}(C, A,\ldots, A)&=&\delta^\ast N^{-1}\mathcal{J}^R_{m+2}(C,C, A,\ldots, A)\,,\\[3mm]
\ell_{m+3}(C, C, A,\ldots, A)&=&\delta^\ast N^{-1}\mathcal{J}^R_{m+3}(C,C,C, A,\ldots, A)
\end{array}
\end{equation}
solve Eqs. (\ref{hpt}). This solution is by no means unique as one can add to the cochains (\ref{rr}) any coboundaries  $\delta \mathcal{L}$ with appropriate numbers of arguments. The corresponding L$_\infty$ algebras are known to be related by a quasi-isomorphism. Notice that the operators $\delta$, $\delta^\ast$, $N$, and $N^{-1}$ respect unitality and the structure maps (\ref{rr}) are unital whenever $\mathcal{J}^R_{m+2}(f,g,A^{\otimes m})$ and $\mathcal{J}^R_{m+3}(f,g,h, A^{\otimes m})$ are so.  
Thus, the recurrence relations (\ref{rr}) allow one to algorithmically extend the de Rham deferential $\ell_1=\dd$ and 
{\it any} unital bidifferntial operator $\ell_2(f,g)$ to an entire L$_\infty$-structure on $\Omega^0(\mathbb{R}^p)\oplus \Omega^1(\mathbb{R}^p)$.

 \subsection{Example}\label{sec:expl}
By way of illustration, let us reconstruct the trilinear map $\ell_3(f,A,A)$ from the generalized Jacobi identity ${\cal J}_3(f,g,A)=0$. Explicitly, 
\begin{eqnarray}\label{b66}
&&\ell_3(\ell_1(f),g,A)+\ell_3(f,\ell_1(g),A)={\cal J}^R_3(f,g,A)\,,\\[3mm]
&&{\cal J}^R_3(f,g,A)=-\ell_1(\ell_3(f,g,A))-\ell_2(\ell_2(f,g),A)-\ell_2(\ell_2(A,f),g)-\ell_2(\ell_2(g,A),f)\,.\notag
\end{eqnarray}
As explained in the previous section, we first replace $f,g\rightarrow C$ and write ${\cal J}^R_3(C,C,A)$ as a polynomial in the formal variables (\ref{zuw}) with coefficients in one-forms:
\begin{equation}
\label{b77}
{\cal J}^R_3(C, C,A)= \big(G_{a\alpha\beta \gamma}z^\alpha z^\beta u^\gamma   + \bar G_{a\alpha\beta J}z^\alpha z^\beta w^J \big)\dd x^a\,.
\end{equation}
The cocycle condition 
\begin{equation}
    (\delta{\cal J}^R_3)(C, C, C)=G_{a\alpha\beta \gamma}z^\alpha z^\beta z^\gamma\dd x^a=0
\end{equation}
implies that the coefficients $G_{a\alpha\beta \gamma}$ have the symmetry of the hook-shaped Young diagram in the indices $\alpha,\beta,\gamma$, that is,  
\begin{equation}
    G_{a\alpha\beta \gamma}=-G_{a\beta\alpha \gamma}  \,,\qquad G_{a\alpha\beta \gamma}+G_{a\beta\gamma \alpha}+  G_{a\gamma\alpha \beta}=0   \,.\label{hook}
    \end{equation}
The operator  $N^{-1}$ applied to  (\ref{b77}) multiplies the first summand by $1/3$ and the second by $1/2$.  Applying then the operator $\delta^\ast$, we finally get 
\begin{equation}
    \ell_3(C,A,A)=-\frac23G_{a\alpha\beta \gamma}z^\alpha u^\beta u^\gamma\dd x^a+\bar G_{a\alpha\beta J}z^\alpha u^\beta w^J \dd x^a
    \end{equation}
or, equivalently, 
\begin{equation}
\begin{array}{c}
\ell_3(f,A,A)=\displaystyle -\frac 23\sum_{p,q,r=1}^\infty {G}_a^{(i)^p(j)^{q}(k)^{r}l}
\left(\partial_i\right)^pf\left(\partial_j\right)^{q-1}A_{j_q}\left(\partial_k\right)^{r-1}  A_{k_{r}}\dd x^a\\[6mm]
\displaystyle -\sum_{p,q,r=1}^\infty {\bar G}_a^{(i)^p(j)^{q}(k)^{r}l}
\left(\partial_i\right)^pf\left(\partial_j\right)^{q-1}A_{j_q}\left(\partial_k\right)^{r-1} (\partial_l A_{k_{r}}-\partial_{k_{r}}A_l)\dd x^a\,.
\end{array}
\end{equation}

\section{Matter fields as an L$_\infty$ module}\label{sec:module}
In this section,  we suggest  a simple way to incorporate  matter fields in the L$_\infty$ approach to noncommutative gauge theories.
In conventional gauge theories, matter fields belong to a representation of the Lie algebra of gauge transformations. Similarly, passing to noncommutative spaces, we assume the matter fields to form an L$_\infty$ module \cite{Lada:1994mn}.

\begin{definition}
Consider an L$_\infty$ algebra $(V,\ell_n)$. Let $M$ be a graded vector space and let $k_n$ be multilinear maps 
\begin{equation}
k_n:V^{\otimes (n-1)}\otimes M\to M .\label{kn}
\end{equation}
of degree $2-n$.
We extend $k_n$ to the elements $v_1,\dots,v_n\in V$ by the equation $k_n(v_1,\dots,v_n):=\ell_n(v_1,\dots,v_n)$. Then $(M,k_n)$ is an L$_\infty$ module if $(V\oplus M,k_n)$ is an L$_\infty$ algebra.
\end{definition}

We consider the case of a single scalar field $\varphi$. The space $M$ has a single homogeneous component. The grading assignment for this component does not play any role. We choose $|\varphi|=1$. The only non-vanishing multilinear maps involving $\varphi$ and compatible with the degree counting are $k_{n+2}(f,A^{\otimes n},\varphi)$. We define the gauge transformation of $\varphi$ as
\begin{eqnarray}
\label{varphi1}
  \delta_{f}  \varphi=-[\![f,\varphi]\!]_A &:=&\sum_{n= 0}^\infty   {1\over n!}
      (-1)^{n(n+1)\over 2}\,
 k_{n+2}(f, A, \dots, A,\varphi )\\&=&k_2(f,\varphi)-k_3(f,A,\varphi)-{1\over2}k_4(f,A,A,\varphi)+\dots\, .\notag
\end{eqnarray} 
It is easy to check that the homotopy relations
\begin{eqnarray}
{\cal J}_{n+3}(f,g,A,\dots,A,\varphi)=0\,,\label{JfgvarphiA}
\end{eqnarray}
imply the closure condition
\begin{align}\label{closurevarphi2}
[\delta_{f},\delta_{g}]\varphi
  = \delta_{[\![f,g]\!]_A}\varphi  
\end{align}
so that the algebra of gauge transformations of $\varphi$ is consistent with the algebra of gauge transformations of $A$ with the bracket $[\![f,g]\!]_A$  given by (\ref{q1}).

Let us outline the construction of lower degree maps $k_n$. Since $M$ has a single component $k_1(\varphi)=0$. For the lowest order gauge variation $\varphi$, we make the choice
\begin{equation}
 \delta_{f}  \varphi=[\![\varphi,f]\!]_A =f\star\varphi+{\cal O}(A)\,, \label{ic3a}
\end{equation}
which is  compatible with (\ref{ic3}) and implies
\begin{equation}
k_2(f,\varphi)=f\star\varphi\,. \label{lp2}
\end{equation}
This choice is motivated by correspondence with U$(1)$ gauge theories on simple noncommutative spaces (e.g. the Moyal plane). The correspondence principle, however, does not allow one to fix $k_2$ uniquely. One can replace (\ref{lp2})  with the right product by the gauge parameter or with the star-commutator. The latter option is less interesting since the corresponding gauge transformation vanishes in the commutative limit. We also do not consider star-anticommutators since the corresponding transformations do not form a closed algebra.

The map $k_3(f,A,\varphi)$ is now determined from the homotopy relation ${\cal J}_3(f,g,\varphi)=0$. This leads to the equation
\begin{eqnarray}\label{p3}
&&k_3(\ell_1(f),g,\varphi)+k_3(f,\ell_1(g),\varphi)={\cal A}_\star(f,g,\varphi)-{\cal A}_\star(g,f,\varphi)\,,
\end{eqnarray}
where the star associator is given by
\begin{equation}\label{assoc}
{\cal A}_\star(f,g,\varphi):=f\star\left(g\star \varphi\right)-\left(f\star g\right)\star \varphi\,.
\end{equation}
Since we are working with a unital star product the associator contains at least one derivative of $f$, $g$, and $\varphi$. Therefore, we can write
\begin{equation}\label{assoc1}
{\cal A}_\star(f,g,\varphi)=\hat{\cal A}_\star(\dd f,\dd g,\dd \varphi)\,.
\end{equation}
It is  easy to check that 
\begin{equation}
k_3(f,A,\varphi)=\frac12\left(\hat{\cal A}_\star(\dd f,A,\dd\varphi)-\hat{\cal A}_\star(A,\dd f,\dd\varphi)\right)
\end{equation}
satisfies  Eq. (\ref{p3}).

With the general technique described in  Section \ref{sec:constr}, it is not hard to write explicit formulas for determining the maps $k_n$. Introducing the ghost field $C$, we can bring the chain of homotopy  relations $\mathcal{J}_{m+2}(f,A^{\otimes m},\varphi)=0$ into the form of homological perturbation theory:
\begin{equation}\label{hpt1}
    (\delta k_{m+2})(C,C; A,\ldots, A;\varphi)=\mathcal{J}_{m+2}^R(C,C;A,\ldots, A; \varphi)\,.
\end{equation}
Here the right-hand side involves $k_{n+2}$ with $n<m$ and the differential $\delta$ essentially coincides with (\ref{delta}):
\begin{equation}
   ( \delta \mathcal{L})(C,\ldots, C; A,\ldots, A;\varphi)= q\mathcal{L}(C,\ldots C; \dd C, A,\ldots, A;\varphi)
   \end{equation}
   for all
   \begin{equation}
\mathcal{L}(\underbrace{C,\ldots C}_{p}; \underbrace{A,\ldots, A}_{q};\varphi)\,.
   \end{equation}
Notice that  $\varphi\in M$ enters these relations as an external parameter and is not affected by $\delta$. 
Again, one can see by induction that $\delta \mathcal{J}^R_{m+2}\equiv 0$ provided  that all  previous equations $\delta k_{n+2}=\mathcal{J}^R_{n+2}$ with $n<m$ are satisfied. Furthermore, by construction, the polydifferential operator $\mathcal{J}^R_{m+2}$ is unital and we may use the same homotopy operator as in (\ref{rr}) to write  down the general solution to (\ref{hpt1}):
\begin{equation}
    k_{m+2}(C; A,\ldots, A;\varphi)=\delta^\ast N^{-1}\mathcal{J}^R_{m+2}(C, C; A,\ldots, A;\varphi)+\mathcal{L}(\dd C, A,\ldots, A;\varphi)\,,
    \end{equation}
$\mathcal{L}$ being an arbitrary polydifferential operator on $\wedge^{m+1}V_1\otimes M$ with values in  $M$.

\section{A$_\infty$ and P$_\infty$ algebras}\label{sec:APinfty}

\begin{definition}\label{def:Ainfty}
A (flat) A$_\infty$ algebra is a $\mathbb{Z}$ graded vector space $V=\bigoplus_{k\in  \mathbb{Z}}\, V_{k}$ together with a system of multilinear maps $
m_n:V^{\otimes n}\to V$, $n\in\mathbb{N}$ of degree $2-n$ satisfying the Stasheff identities
\begin{equation}
\sum_{\substack{ r+s=n+1\\ 1\leq i \leq r}}(-1)^{\epsilon(m,r,s,i)}m_r(v_1,\dots,v_{i-1},m_s(v_i,\dots,v_{i+s-1}),v_{i+s},\dots,v_n)=0\,,\label{Ainf}
\end{equation}
where
\begin{equation}
\epsilon(n,r,s,i)=(s+1)i+sn+s(|v_1|+\dots+|v_{i-1}|)
\end{equation}
for $n\in\mathbb{N}$.
\end{definition}
We took the sign conventions from \cite{Markl:2012}.  Flatness means the absence of $m_0$ map, i.e. $m_0=0$. By writing the Stasheff identity (\ref{Ainf}) involving $m_2(f,m_2(g,h))$ one can see that the binary product $m_2(f,g)$ is associative up to homotopy.

Let $(V,m_n)$ be an A$_\infty$ algebra. Define
\begin{equation}
\ell_n(v_1,\dots,v_n)=\sum_{\sigma\in S_n} \chi(\sigma;v_1,\dots, v_n) m_n(v_{\sigma(1)},\dots,v_{\sigma(n)}).\label{mell}
\end{equation}
Then, $(V,\ell_n)$ is an L$_\infty$ algebra. Thus, A$_\infty$ algebras give L$_\infty$ algebras in a way much similar to the one in which associative algebras give Lie algebras.

The usual approach \cite{Bayen:1977ha} to deformation quantization starts with a commutative algebra $C^\infty(M)$ of smooth functions on some manifold $M$ with the point-wise product $f,g\to f\cdot g$ and a Poisson bracket $\{ \cdot\,,\cdot \}$ which is a derivation on this algebra, $\{ f,g\cdot h\}=\{ f,g\}\cdot h + g\cdot \{f,h\}$. It is also assumed that the bracket satisfies the Jacobi identity, so that $(C^\infty(M),\{\cdot\, , \cdot\})$ becomes a Lie algebra. The existence of deformation quantization and an explicit construction follow from the Kontsevich formality theorem. 

In the present work we deal with a graded vector space $V$ (rather than $C^\infty(M)$). Thus let us fix a graded commutative product on $V$, i.e., a map $\mu: V\otimes V \to V$. A degree $k$ derivation $D$ on $V$ is a linear map such that (i) $DV^j\subset V^{j+k}$ and (ii) $D \mu (u,v)=\mu (Du,v)+(-1)^{k|u|}\mu(u,Dv)$.

Another ingredient, namely, a strong homotopy Poisson algebra (or a P$_\infty$ algebra) was defined by Cattaneo and Felder in \cite{Cattaneo:2007}, see also \cite{cVORONOV2005133}, \cite{Lyakhovich_2005}. 
\begin{definition}\label{def:Pinfty}
A flat P$_\infty$ algebra is a flat L$_\infty$ algebra defined on a $\mathbb{Z}$-graded vector space $V=\bigoplus_{k\in  \mathbb{Z}}\, V_{k}$ with a graded commutative product $\mu$ and multilinear maps  $p_n$, $n\geq 1$,  such that the maps
\begin{equation}
v\to p_n(v_1,\dots,v_{n-1},v)
\end{equation}
are derivations on $V$ of degree $2-n+\sum_{i=1}^{n-1}|v_i|$.
\end{definition}

Let $V^{(0)}$ be a graded vector space, and let $V=V^{(0)}[[\lambda]]$. Consider an A$_\infty$ algebra on $V$. The multilinear maps $m_n$ can be represented as formal power series in $\lambda$, $m_n=m_n^{(0)}+\lambda m_n^{(1)}+\dots$. Let us assume that $m_n^{(0)}=0$ for $n\neq 2$ and $m_2^{(0)}$ is a graded commutative multiplication on $V^{(0)}$. Define $p_n=\sum_{\sigma\in S_n} \chi(\sigma)\, m_n^{(1)} \circ \sigma$. It was demonstrated in \cite{Cattaneo:2007} that $(V^{(0)},p_n)$ is a P$_\infty$ algebra with $\mu=m_2^{(0)}$ This means, that quasiclassical limits of some formal A$_\infty$ algebras are P$_\infty$ algebras.

This result allows us to derive some no-go statements. Consider a graded vector space of formal power series of 0-forms (in $V_0$) and 1-forms (in $V_1$) on $\mathbb{R}^N$, but now we do not interpret the 0-forms as gauge parameters. We are interested in an A$_\infty$ algebra which in the classical limit $\lambda\to 0$ describes just the 0-forms, i.e. the only non-vanishing $m_n^{(0)}$ is $m_2^{(0)}(f,g)=f\cdot g$. Thus, the condition from the previous paragraph is satisfied and we may try to construct a P$_\infty$ algebra. We additionally assume that 
\begin{equation}
p_2(f,g)=\{ f,g\}, \label{p2}
\end{equation}
where $\{ \cdot\, , \cdot\}$ is an almost Poisson bracket. Since $p_1$ is a derivation of $m_2^{(0)}$, we have 
\begin{equation*}
p_1(f\cdot g)=m_2^{(0)}(p_1 f,g)+m_2^{(0)}( f,p_1g).
\end{equation*}
The right hand side of this equation vanishes since $m_2^{(0)}( f,A)$ for any $A$. By setting $g=1$ we obtain that $p_1$ vanishes identically. The equation ${\cal J}_3(f,g,h)=0$ yields
\begin{equation*}
p_2(p_2(f,g),h)+p_2(p_2(g,h),f)+p_2(p_2(h,f),g)=0,
\end{equation*}
i.e. the almost Poisson bracket  $\{ \cdot ,\cdot\}$ has to satisfy the Jacobi identity.

The paper \cite{Cattaneo:2007} (especially the Relative Formality Theorem) suggests that P$_\infty$ algebras are natural objects in the context of deformation quantization. Therefore, it makes sense to study other P$_\infty$ algebras even if they do not fit into the construction of quasiclassical limits of A$_\infty$ algebras described above. Our next example operates with the same graded vector space $V$ and uses the same ansatz (\ref{p2}) for $p_2(f,g)$. The product $\mu$ is taken to be the usual product on the truncated de Rham complex. Namely,
\begin{equation}
\mu(f,g)=f\cdot g,\qquad \mu(f,A)=\mu(A,f)=f\cdot A.
\end{equation}
We take also $p_1(f)=\dd f$. No further conditions are imposed. This example is closer to our construction of L$_\infty$ algebras in the preceding sections.

 The Leibniz rule for $p_2 (f,\cdot)$ yields
\begin{equation}
p_2(f,gA)=gp_2(f,A)+p_2(f,g)A.\label{diffell2}
\end{equation}
Thus, $p_2 (f,A)$ can be represented as a sum of first-order and zeroth-order differential operators
\begin{eqnarray}
p_2(f,A)_k=\mathcal{L}_{2,0}(f)_{k}^iA_i + \mathcal{L}_{2,1}(f)_{k}^{ij}\partial_iA_j\,.
\end{eqnarray}
The condition (\ref{diffell2}) does not restrict $\mathcal{L}_{2,0}$ but fixes 
\begin{equation}
\mathcal{L}_{2,1}(f)_{k}^{ij}=\delta_k^j (\partial_l f)P^{li},
\end{equation}
where $P^{li}$ is the bivector defining the almost Poisson bracket in (\ref{p2}).
A further $L_\infty$ condition reads
\begin{equation}
\dd \{ f, g\} =p_2(\dd f,g)+p_2(f,\dd g).
\end{equation}
The general solution to this equation is
\begin{equation}
\mathcal{L}_{2,0}(f)^i_k=-\tfrac 12\partial_k P^{ij}\partial_j f +Q^{ij}_k\partial_j f\,,
\end{equation}
where $Q_k^{ij}$ is any $x$-dependent function symmetric with respect to $i\leftrightarrow j$. By equating the terms containing the same numbers of derivatives of independent fields in conditions $\mathcal{J}_3(f,g,h)=0$ and in the derivative conditions for $p_3(f,g,A)$ one can easily find a general solution
\begin{equation}
p_3(f,g,A)=\left( \tfrac 13 \hat J_3^{ijk} +Q'^{\ ijk}\right)\partial_i f\cdot \partial_j g\cdot A_k,
\end{equation}
where
\begin{equation}
\hat J_3^{ijk} =P^{il}\partial_l P^{jk} +P^{jl}\partial_l P^{ki}+P^{kl}\partial_l P^{ij}
\end{equation}
is the jacobiator and $Q'^{\ ijk}$ is any tensor with the symmetry of the hook-shaped Young diagram, see Eq. (\ref{hook}). 

Looking at the way $Q$ appears in the formulas one concludes that it should be linear in $P$. The only available tensor structure of this type is $\partial_k P^{ij}$ but it does not have required symmetry in the upper indices.  Thus, in what follows we set $Q=0$.

Consider the he derivation condition
\begin{equation}
p_3(f,A,gB)=g p_3(f,A,B)+p_3(f,A,g)B.\label{der3}
\end{equation} 
Obviously, $p_3$ can only be a zeroth or first order differential operator in the last argument,
\begin{equation}
p_3(f,A,B)_k=\mathcal{L}_{3,0}(f,A)_k^iB_i+\mathcal{L}_{3,1}(f,A)_k^{ij}\partial_iB_j\,.
\end{equation}
By substituting this form of $p_3$ in the equation (\ref{der3}) and comparing the terms having the same number of derivatives of independent field we obtain
\begin{equation}
\mathcal{L}_{3,1}(f,A)_k^{ij}=\left[ \tfrac 13 \delta_k^j\hat J^{ipl} +Q'^{ipl} \right] \partial_pf\cdot A_l.
\end{equation}
$\mathcal{L}_{3,0}$ is not restricted by the condition (\ref{der3}).
Let us take the equation
\begin{eqnarray}
&&0=\mathcal{J}_3(f,g,A)=p_1(p_3(f,g,A))+p_3(\dd f,g,A)+p_3(f,\dd g,A)\nonumber\\
&&\qquad\qquad +p_2(p_2(f,g),A)+p_2(p_2(A,f),g)+p_2(p_2(g,A),f),
\end{eqnarray}
fix a point $x$, and choose $A$ such that $A(x)=0$. Then, at this point,
\begin{equation}
0=\left[ \tfrac 13 \hat J^{ijl}-Q'^{\ lij}-Q'^{\ lji} \right] \partial_i f\cdot\partial_j g \cdot (\partial_k A_l -\partial_l A_k)\label{Jac0}
\end{equation}
Since $x$ is arbitrary, and the derivatives of $A$ are arbitrary at this point, we have the condition
\begin{equation}
    \tfrac 13 \hat J^{ijl}-Q'^{\ lij}-Q'^{\ lji} =0.
\end{equation}
After taking a sum over cyclic permutations of $(i,j,l)$ we obtain $\hat J^{ijl}=0$, i.e., 
the jacobiator has to vanish. 

We are led to conclude that P$_\infty$ implies that the Poisson structure is Lie--Poisson.

Let us summarize the results of this section. We have considered two possible choices for the graded commutative product for two-term P$_\infty$ algebras with $p_2(f,g)=\{ f,g\}$ given by some almost Poisson bracket. For both choices, the Jacobi identity on this bracket necessarily follows from the P$_\infty$ relations. In that case, the construction of the corresponding P$_\infty$ algebra was given in \cite{Kupriyanov:2021cws}. We choose the graded commutative product $\mu$ being an undeformed pointwise product on some truncation of the de Rham complex. Other choices for $\mu$ may  change our conclusions regarding the existence of the P$_\infty$ algebra.

\section{Conclusion}\label{sec:concl}
Let us briefly summarize  the main results of the paper. We proved the existence and derived inductive formulas for the structure maps of an L$_\infty$ algebra describing U$_\star(1)$ gauge transformations on an arbitrary noncommutative (and even nonassociative) space. The same is done with an L$_\infty$ module describing matter fields. We also attempted to include P$_\infty$ algebras in this approach and ended up in two no-go statements.

Our approach is rather general and the formulas obtained are very simple. In fact, due to the use of proper mathematical machinery, they look even simpler than particular results for a few lower order $\ell_n$ existing in the literature \cite{Blumenhagen:2018kwq,Kupriyanov:2021cws, KupDurham}. (Of course, after bringing our formulas to an expanded component form they are equivalent.) This suggests that the method will be efficient for studying various extensions of the scheme considered here. One possible  extension is the inclusion of a nontrivial $V_2$ space containing the field strength two-forms \cite{Hohm:2017pnh,Kupriyanov:2019cug} and hence dynamics. Another extension is to noncommutative deformations of non-abelian gauge theories, which may help to overcome the restrictions found in \cite{Chaichian:2001mu}. We also note that a solution to this problem has been suggested basing on braided noncommutativity \cite{DimitrijevicCiric:2021jea}, see \cite{Giotopoulos:2021ieg} for a review.

\paragraph{Acknowledgements.} This work was supported in parts by the S\~ao Paulo Research Foundation (FAPESP), grants 2021/09313-8 (V.K.), 2021/10128-0 (D.V.) and 2022/13596-8 (A.Sh),  and by the National Council for Scientific and Technological Development (CNPq), grants 304130/2021-4 (V.K.) and 304758/2022-1 (D.V.). The results of Sec. 3.2 on homological perturbation theory were obtained under the exclusive support of the Ministry of Science and Higher Education of the Russian Federation (project No. FSWM-2020-0033).



\end{document}